\documentclass[aps,prc,twocolumn,groupaddress,showpacs,nofootinbib,floatfix]{revtex4}

\usepackage{bm}
\usepackage{epsfig}
\usepackage{longtable}
\usepackage{dcolumn}

\def\bea {\begin{eqnarray}}
\def\eea {\end{eqnarray}}
\def\be {\begin{equation}}
\def\ee {\end{equation}}
\def\ben{\begin{enumerate}}
\def\een{\end{enumerate}}
\def\bi{\begin{itemize}}
\def\ei{\end{itemize}}

\def\prl {Phys. Rev. Lett.\ }

\def\pr {Phys. Rev.\ }

\newcommand{\sfrac}[2]{\mbox{\small{$\frac{#1}{#2}$}}}

\begin{document} 

\title{Parameterization of the statistical rate function for select superallowed transitions}

\author{I.S. Towner}
\email{towner@comp.tamu.edu}
\author{J.C. Hardy} 
\email{hardy@comp.tamu.edu}
\affiliation{Cyclotron Institute, Texas A\&M University, College Station, Texas 77843}

\date{\today}

\begin{abstract}
We present a parameterization of the statistical rate function, $f$, for 20 superallowed $0^+$$\rightarrow 0^+$ nuclear
$\beta$ transitions between $T$=1 analog states, and for 18 superallowed ``mirror'' transitions between analog $T$=1/2 states.  All
these transitions are of interest in the determination of $V_{ud}$.  Although most of the transition $Q_{EC}$ values have been
measured, their precision will undoubtedly be improved in future.  Our parameterization allows a user to easily calculate the
corresponding new $f$ value to high precision ($\pm$0.01\%) without complicated computing.
\end{abstract}


\maketitle

\section{Introduction}

Precise measurements of nuclear $\beta$ decay provide a valuable window into the electroweak standard model.  In particular,
superallowed $0^+$$\rightarrow 0^+$ transitions between $T$=1 analog states are used to set a limit on the presence of
scalar interactions and to determine $V_{ud}$, the upper left element of the Cabibbo-Kobayashi-Maskawa (CKM) matrix and a
key contributor to the most demanding available test of the unitarity of that matrix.  While these transitions currently lead
to the most precise determination of $V_{ud}$, mirror transitions between $T$=$\,$1/2 analog states are becoming of interest
as a means of confirming $V_{ud}$ via a different experimental approach.  To be useful, not only must the $Q_{EC}$ value for
each of these transitions be measured very precisely but the statistical rate function, $f$, which uses the $Q_{EC}$ as input, 
must be calculated with equivalent precision. 

Because there is no widely available means for calculating $f$ to the required level of precision, we have devised a simple
parameterization that reproduces the results of our full code for energies spanning a small range around the currently known
$Q_{EC}$ values for both types of superallowed transition.  Together, these should provide a convenient resource for
experimentalists to use in future to obtain high-precision $f$ values from improved $Q_{EC}$-value measurements for
these transitions.

Our goal in what follows is to parameterize $f$ and present tables of the parameters for the two sets of transitions: 1) the
20 superallowed $0^+$$\rightarrow 0^+$ nuclear $\beta$ transitions between $T$=1 analog states, whose properties have been
surveyed in Refs.~\cite{HT09, HT15}; and 2) the 18 superallowed ``mirror'' $\beta$ transitions between the analog $T$=1/2 states surveyed
in Ref.~\cite{Se08}.  For each transition, we have computed $f$ for 100 values of $Q_{EC}$ taken over a range of $\pm 60$ keV
around the transition $Q_{EC}$-value\footnote{For $^{70}$Br the $Q_{EC}$-value is less precisely known, so the $Q_{EC}$-value range for fitting
was extended to $\pm 600$ keV.} and fitted these results to determine the coefficients in our parameterization.  Our aim in fitting
these 100 values is to achieve an accuracy of $0.01 \%$, nearly a factor of ten more precise than is currently required.  

\section{Parameterization of the statistical rate function}
\label{s:fexact}

To achieve 0.01\% accuracy, the electron wave function must be determined with great precision.  In our detailed evaluation of $f$
\cite{HT05}, we accomplished this by solving the Dirac equation for the emerging electron moving in the Coulomb field of the nuclear
charge distribution.  The full expression for the computation of $f$ is
\bea
f & = & \xi R(W_0) \int_1^{W_0} p W (W_0-W)^2 F(Z,W) f_1(W)
\nonumber \\
 & & ~~~~~~~~~~~~~~~~~~~~~~~Q(Z,W) r(Z,W) ~dW ,
\label{fexact}
\eea
where $W$ is the electron total energy in electron rest-mass units, $W_0$ is the maximum value of $W$, $p = 
( W^2-1)^{1/2}$ is the electron momentum, $Z$ is the charge number of the daughter nucleus (positive for electron
emission, negative for positron emission), $F(Z,W)$ is the Fermi function and $f_1(W)$ is the shape-correction
function as defined by Holstein \cite{Hol74} (but with kinematic recoil corrections omitted).

Further, $Q(Z,W)$ is a screening correction for which we use the analytic prescription of Rose \cite{Ro36} (see
Eq.~(A44) in Ref.~\cite{HT05}), and $r(Z,W)$ is an atomic overlap correction described in Ref.~\cite{HT09}.  The kinematic
recoil corrections that Holstein includes in $f_1(W)$ are here written as $R(W_0)$.  The expression for $R(W_0)$
is derived in Appendix~\ref{s:recoil}, with the result that
\be
R(W_0) \simeq 1 - \frac{3 W_0}{2 M_A} ,
\label{recoil}
\ee
where $M_A$ is the average of the initial and final nuclear masses expressed in electron-mass units.  Last, for allowed
transitions it is customary to remove the leading nuclear matrix element from the definition of $f$.  Thus we have
introduced $\xi$ in Eq.~(\ref{fexact}), where $\xi = 1/|{\cal M}_F|^2$ for superallowed Fermi transitions, ${\cal M}_F$
being the Fermi matrix element.  For mixed Fermi and Gamow-Teller transitions, 
$\xi = 1/[{\cal M}_F^2 + g_A^2 {\cal M}_{GT}^2 ]$ with ${\cal M}_{GT}$
being the Gamow-Teller matrix element and $g_A$ the axial-vector coupling constant.

\begin{table*}
\caption{Values of the coefficients $a_0$ and $a_1$ that yield the statistical rate function $f_0$ from Eq.~(\ref{f0fit}), and
coefficients $b_0$, $b_1$, $b_2$ and $b_3$ that yield the correction $\delta_S$ from Eq.~(\ref{ds2}).  Coefficients $a_2$ and
$a_3$ are held fixed at the values: $a_2 = -2/15$ and $a_3 = 1/4$.  The cases shown are the superallowed Fermi transitions between $T$=1,
$J^{\pi}$=$\,$$0^+$ analog states surveyed in Ref.~\cite{HT09, HT15}.
\label{t:tab1}}
\begin{ruledtabular}
\begin{tabular}{rrrrrrr}
& & & & & & \\[-3mm]
\multicolumn{1}{r}{Parent} & & & & & & \\
\multicolumn{1}{r}{nucleus}
 & \multicolumn{1}{c}{$a_0$}
 & \multicolumn{1}{c}{$a_1$}
 & \multicolumn{1}{c}{$b_0(\%)$}
 & \multicolumn{1}{c}{$b_1(\%)$}
 & \multicolumn{1}{c}{$b_2(\%)$}
 & \multicolumn{1}{c}{$b_3(\%)$}
 \\[1mm]
\hline
& & & & & & \\[-3mm]
   $^{10}$C &                                     
   0.0297225  &  $-0.1431540$  &
   0.01178  &   0.02006  &   0.05203  &  $-0.00096$ \\
   $^{14}$O &                                     
   0.0285463  &  $-0.1417222$  &
   0.03176  &   0.03123  &   0.06506  &  $-0.00101$ \\
   $^{18}$Ne  &                                    
   0.0274005  &  $-0.1398743$  &
   0.04750  &   0.04995  &   0.08945  &  $-0.00134$ \\
   $^{22}$Mg  &                                   
   0.0263237  &  $-0.1374785$  &
   0.07036  &   0.06393  &   0.10796  &  $-0.00139$ \\
   $^{26}$Si  &                                   
   0.0253385  &  $-0.1369946$  &
   0.10306  &   0.07827  &   0.12856  &  $-0.00146$ \\
   $^{30}$S  &                                    
   0.0242904  &  $-0.1271838$  &
   0.14905  &   0.09529  &   0.15413  &  $-0.00161$ \\
   $^{34}$Ar  &                                   
   0.0233252  &  $-0.1182701$  &
   0.15336  &   0.11896  &   0.18478  &  $-0.00184$ \\
   $^{38}$Ca  &                                   
   0.0223867  &  $-0.1018182$  &
   0.17301  &   0.13558  &   0.20674  &  $-0.00186$ \\
   $^{42}$Ti  &                                   
   0.0216593  &  $-0.1105386$  &
   0.15625  &   0.15293  &   0.23380  &  $-0.00196$ \\[2mm]
   $^{26m}$Al  &                               
   0.0257927  &  $-0.1355697$  &
   0.09813  &   0.07208  &   0.12037  &  $-0.00149$ \\
   $^{34}$Cl  &                                   
   0.0238533  &  $-0.1281700$  &
   0.16759  &   0.10598  &   0.16911  &  $-0.00173$ \\
   $^{38m}$K  &                                
   0.0228360  &  $-0.1090747$  &
   0.17630  &   0.12480  &   0.19325  &  $-0.00180$ \\
   $^{42}$Sc  &                                   
   0.0220302  &  $-0.1082192$  &
   0.17335  &   0.14265  &   0.22147  &  $-0.00193$ \\
   $^{46}$V  &                                     
   0.0211437  &  $-0.0894977$  &
   0.20200  &   0.16556  &   0.25213  &  $-0.00208$ \\
   $^{50}$Mn  &                                   
   0.0202722  &  $-0.0597791$  &
   0.25281  &   0.18330  &   0.27834  &  $-0.00214$ \\
   $^{54}$Co  &                                   
   0.0195698  &  $-0.0524836$  &
   0.32812  &   0.19757  &   0.30337  &  $-0.00217$ \\
   $^{62}$Ga  &                                   
   0.0181322  &  $-0.0141676$  &
   0.49342  &   0.24418  &   0.36915  &  $-0.00243$ \\
   $^{66}$As  &                                   
   0.0173202  &   0.0473840  &
   0.57218  &   0.27260  &   0.40770  &  $-0.00261$ \\
   $^{70}$Br  &                                   
   0.0167829  &   0.0417070  &
   0.56671  &   0.29670  &   0.43602  &  $-0.00265$ \\
   $^{74}$Rb  &                                   
   0.0162385  &   0.0446304  &
   0.44643  &   0.32180  &   0.46490  &  $-0.00269$ \\[1mm]
\end{tabular}
\end{ruledtabular}
\end{table*}

In order to parameterize $f$, it is convenient to factor it into two contributions:
\be
f = f_0 (1 + \delta_S) ,
\label{ffactor}
\ee
\be
f_0 = \int_1^{W_0} p W (W_0 - W)^2 F(Z,W) Q(Z,W) r(Z,W) ~dW ,
\label{f0}
\ee
\be
\delta_S = (f - f_0)/f_0 .
\label{dels}
\ee
The purpose of this factorization is to place the role of the shape-correction function $f_1(W)$ entirely within the
correction term $\delta_S$, which is typically of the order of a few percent.  The shape-correction function depends
on nuclear matrix elements and differs for Fermi and Gamow-Teller transitions.  This piece of the calculation is
somewhat less certain since it is nuclear-structure dependent; however, being small, its accuracy is also less critical.

In the limit that $F(Z,W) Q(Z,W) r(Z,W) \rightarrow 1$, which occurs when $Z=0$, the integral $f_0$ has an analytic
value:
\be
f_0(Z=0) = \sfrac{1}{30} W_0^4 p_0 - \sfrac{3}{20} W_0^2 p_0
- \sfrac{2}{15} p_0 + \sfrac{1}{4} W_0 \ln (W_0 + p_0) ,
\label{f00}
\ee
with $p_0 = (W_0^2 - 1)^{1/2}$.  This suggests a fitting function of the form
\be
f_0 = a_0 W_0^4 p_0 + a_1 W_0^2 p_0 + a_2 p_0 + a_3 W_0 \ln (W_0 + p_0) .
\label{f0fit}
\ee
In fitting 100 values of $f_0$, we found that the four parameters $a_0, a_1, a_2$ and $a_3$ could not be uniquely determined
with precision.  Thus it was decided to fix $a_2$ and $a_3$ to their $Z=0$ values, namely $a_2 = -2/15$ and $a_3 = 1/4$, and
use the fitting process to determine $a_0$ and $a_1$.  This procedure yielded the required precision for $f_0$.  The
resultant values of $a_0$ and $a_1$ are given in Table~\ref{t:tab1} for the $0^+$$\rightarrow 0^+$ transitions, and in
Table~\ref{t:tab2} for the $T$=$\,$1/2 mirror transitions.

For the correction $\delta_S$ we start with the approximate expression
\bea
f_0 \delta_S & \simeq & \int_1^{W_0} p W (W_0 - W)^2 F(Z,W)
\nonumber \\
& & ~~~~~~~~~~~~ \left [ \xi f_1(W) - 1 - \frac{3 W_0}{2 M_A} \right ] ~ dW
\label{f0ds}
\eea
and write
\be
\xi f_1(W) - 1 = B_0 +B_1 W + B_2/W + B_3 W^2 .
\label{f1}
\ee
The coefficients $B_0$, $B_1$, $B_2$ and $B_3$ are different for Fermi and Gamow-Teller transitions.
This choice of parameterization is guided by the early work of Schopper
\cite{Sc66}
who used such a parameterization for the shape-correction function.

\subsection{Superallowed $0^+$$\rightarrow 0^+$ Fermi transitions}
\label{ss:Fermi}

For Fermi (vector) transitions,
\bea
B_0^F & = & -\sfrac{1}{5} (W_0R)^2 + \sfrac{1}{15} R^2 - \sfrac{6}{35}
(\alpha Z) (W_0 R) + \sfrac{61}{630}(\alpha Z)^2 ,
\nonumber \\
B_1^F & = & \sfrac{4}{15} (W_0 R) R - \sfrac{48}{35} (\alpha Z) R ,
\nonumber \\
B_2^F & = & \sfrac{2}{15} (W_0 R) R - \sfrac{18}{35} (\alpha Z) R ,
\nonumber \\
B_3^F & = & - \sfrac{4}{15} R^2 ,
\label{BnF}
\eea
where $R$ is the radius of the nuclear charge distribution expressed in electron Compton wavelength units.  We derived these equations from the work
of Behrens and B\"{u}hring \cite{BB82} who give algebraic expressions
for the shape-correction function as expansions in the small quantities
$R$ and $(\alpha Z)$.  Our Eqs.~(\ref{BnF}) and (\ref{BnGT}) below are
correct to second order in these quantities, namely to order
$R^2$, $(\alpha Z)^2$ and $(\alpha Z)R$.  Inserting
Eqs.~(\ref{BnF}) and (\ref{f1}) into Eq.~(\ref{f0ds}) we obtain
\be
\delta_S \simeq B_0 + B_1 \langle W \rangle  + B_2 \langle 1/W \rangle 
+ B_3 \langle W^2 \rangle - \frac{3W_0}{2 M_A} ,
\label{ds1}
\ee
where $\langle W^n \rangle$ is the value of $W^n$ averaged over the electron spectrum.  Estimates of these quantities are:
$\langle W \rangle = W_0/2$, $\langle W^{-1} \rangle = 5W_0^{-1}/2$ and $\langle W^2 \rangle = 2W_0^2/7$.

This leads to our final choice of parameterization for the correction $\delta_S$:
\be
\delta_S = b_0 + b_1 W_0 + b_2/W_0 + b_3 W_0^2 ,
\label{ds2}
\ee
where approximate values of the coefficients are
\bea
b_0^F & \simeq & \sfrac{2}{5} R^2 + \sfrac{61}{630} (\alpha Z)^2 ,
\nonumber \\
b_1^F & \simeq & - \sfrac{6}{7} (\alpha Z) R - \sfrac{3}{2 M_A} ,
\nonumber \\
b_2^F & \simeq & - \sfrac{9}{7} (\alpha Z) R ,
\nonumber \\
b_3^F & \simeq & - \sfrac{1}{7} R^2 .
\label{bbnF}
\eea
We fitted the expression in Eq.~(\ref{ds2}) to the exactly computed value of $\delta_S$ from Eq.~(\ref{dels}) to obtain the
parameters $b_0$, $b_1$, $b_2$ and $b_3$.  Again, it was found that all four parameters could not be uniquely determined with
precision, so the coefficients $b_2$ and $b_3$ were fixed at the values given in Eq.~(\ref{bbnF}) for $b_2^F$ and $b_3^F$, and
the fitting process was used to determine $b_0$ and $b_1$.  Table~\ref{t:tab1} gives the values of the parameters $b_0$, $b_1$,
$b_2$ and $b_3$ for the superallowed $0^+$$\rightarrow 0^+$ Fermi transitions.

\begin{table*}
\caption{Values of the coefficients, $a_0$ and $a_1$, that yield the statistical rate function $f_0$ from
Eq.~(\ref{f0fit}), and coefficients, $b_0$, $b_1$, $b_2$ and $b_3$, that yield the correction, $\delta_S$, from
Eq.~(\ref{ds2}).  For each case, the $b$ coefficients in the first row correspond to the Fermi component, $b_0^F$
etc., and those in the second row correspond to the Gamow-Teller component, $b_0^{GT}$ etc.  Coefficients $a_2$
and $a_3$ are held fixed at values: $a_2 = -2/15$ and $a_3 = 1/4$.  The cases shown are the mixed Fermi and Gamow-Teller 
transitions between mirror $T$=$\,$1/2 states in odd-mass nuclei surveyed in Ref.~\cite{Se08}. 
\label{t:tab2}}
\begin{ruledtabular}
\begin{tabular}{rrrrrrr}
& & & & & & \\[-3mm]
\multicolumn{1}{r}{Parent} & & & & & & \\
\multicolumn{1}{r}{nucleus}
 & \multicolumn{1}{c}{$a_0$}
 & \multicolumn{1}{c}{$a_1$}
 & \multicolumn{1}{c}{$b_0(\%)$}
 & \multicolumn{1}{c}{$b_1(\%)$}
 & \multicolumn{1}{c}{$b_2(\%)$}
 & \multicolumn{1}{c}{$b_3(\%)$}
 \\[1mm]
\hline
& & & & & & \\[-3mm]
   $^{11}$C  &                                    
   0.0297280  &  $-0.1431964$  &
   0.01227  &   0.02003  &   0.05030  &  $-0.00093$ \\
& & &   0.71578  &   0.07695  &   0.56495  &  $-0.00055$ \\
   $^{13}$N  &                                    
   0.0291054  &  $-0.1420911$  &
   0.02166  &   0.02566  &   0.05640  &  $-0.00095$ \\
& & &   0.31915  &   0.04052  &   0.36491  &  $-0.00026$ \\
   $^{15}$N  &                                    
   0.0285370  &  $-0.1416159$  &
   0.04552  &   0.03330  &   0.06846  &  $-0.00117$ \\
& & &   0.25027  &   0.02949  &   0.31261  &  $-0.00024$ \\
   $^{17}$F  &                                    
   0.0279301  &  $-0.1400527$  &
   0.02528  &   0.04132  &   0.07482  &  $-0.00113$ \\
& & &   1.47963  &   0.06556  &   0.67635  &  $-0.00074$ \\
   $^{19}$Ne  &                                   
   0.0273984  &  $-0.1398186$  &
   0.05521  &   0.04998  &   0.08914  &  $-0.00135$ \\
& & &   1.34270  &   0.05876  &   0.55039  &  $-0.00099$ \\
   $^{21}$Na  &                                   
   0.0268709  &  $-0.1394429$  &
   0.05922  &   0.05859  &   0.09992  &  $-0.00142$ \\
& & &   1.73365  &   0.06313  &   0.64005  &  $-0.00101$ \\
   $^{23}$Mg  &                                   
   0.0263324  &  $-0.1378844$  &
   0.06375  &   0.06436  &   0.10731  &  $-0.00138$ \\
& & &   1.91334  &   0.06348  &   0.64697  &  $-0.00093$ \\
   $^{25}$Al  &                                   
   0.0258123  &  $-0.1364357$  &
   0.08154  &   0.07160  &   0.11849  &  $-0.00144$ \\
& & &   2.34766  &   0.07237  &   0.73737  &  $-0.00090$ \\
   $^{27}$Si  &                                   
   0.0252815  &  $-0.1330446$  &
   0.09398  &   0.07980  &   0.12945  &  $-0.00149$ \\
& & &   2.75164  &   0.07192  &   0.79082  &  $-0.00086$ \\
   $^{29}$P  &                                    
   0.0247788  &  $-0.1301483$  &
   0.11110  &   0.08717  &   0.13982  &  $-0.00151$ \\
& & &   2.41055  &   0.06857  &   0.66134  &  $-0.00086$ \\
   $^{31}$S  &                                    
   0.0243506  &  $-0.1328386$  &
   0.14554  &   0.09612  &   0.15460  &  $-0.00163$ \\
& & &   2.17479  &   0.07799  &   0.60883  &  $-0.00092$ \\
   $^{33}$Cl  &                                   
   0.0239077  &  $-0.1333333$  &
   0.15209  &   0.10627  &   0.16772  &  $-0.00170$ \\
& & &  $-0.78012$  &   0.06397  &   0.14681  &  $-0.00050$ \\
   $^{35}$Ar  &                                   
   0.0233631  &  $-0.1226819$  &
   0.19658  &   0.11636  &   0.18453  &  $-0.00183$ \\
& & &  $-0.53603$  &   0.07047  &   0.19515  &  $-0.00055$ \\
   $^{37}$K  &                                    
   0.0229710  &  $-0.1251564$  &
   0.18369  &   0.12491  &   0.19342  &  $-0.00180$ \\
& & &   0.89286  &   0.09010  &   0.41192  &  $-0.00084$ \\
   $^{39}$Ca  &                                   
   0.0224606  &  $-0.1123165$  &
   0.21779  &   0.13259  &   0.20653  &  $-0.00185$ \\
& & &   0.58246  &   0.09757  &   0.37316  &  $-0.00087$ \\
   $^{41}$Sc  &                                   
   0.0220044  &  $-0.1046436$  &
   0.20989  &   0.14001  &   0.22165  &  $-0.00193$ \\
& & &   3.94590  &   0.12635  &   0.91561  &  $-0.00124$ \\
   $^{43}$Ti  &                                   
   0.0216749  &  $-0.1134605$  &
   0.15801  &   0.15650  &   0.23644  &  $-0.00200$ \\
& & &   3.54635  &   0.13277  &   0.82630  &  $-0.00135$ \\
   $^{45}$V  &                                    
   0.0211420  &  $-0.0910683$  &
   0.31418  &   0.16234  &   0.25779  &  $-0.00218$ \\
& & &   4.54443  &   0.14448  &   0.99222  &  $-0.00134$ \\[1mm]
\end{tabular}
\end{ruledtabular}
\end{table*}

\subsection{Mirror $T$=$\,$1/2 transitions}
\label{ss: GT}

For pure Gamow-Teller (axial-vector) transitions, coefficients in the expression for the shape-correction function in Eq.~(\ref{f1})
are: 
\bea
B_0^{GT} & \simeq & - \sfrac{1}{5} (W_0 R)^2 + \sfrac{11}{45} R^2
\left ( 1 - \sfrac{2}{11} x \right )  
\nonumber \\
& & + \sfrac{2}{35} (\alpha Z)
(W_0 R) ( 1 - x)
+ \sfrac{1}{3} (W_0 R) \left [ \mp 2 \overline{b} + \overline{d} \right ]
\nonumber \\
& & + \sfrac{1}{3} \beta (\alpha Z) \left [ \pm 2 \overline{b} +
\overline{d} \right ] + \sfrac{61}{630} (\alpha Z)^2 ,
\nonumber \\
B_1^{GT} & \simeq & \sfrac{4}{9} (W_0 R) R \left ( 1 - \sfrac{1}{10} x \right )
- \sfrac{8}{5} (\alpha Z) R \left ( 1 - \sfrac{1}{28} x \right )
\nonumber \\
& & \pm \sfrac{4}{3} R \overline{b} ,
\nonumber \\
B_2^{GT} & \simeq & - \sfrac{2}{45}(W_0 R)R (1-x) - \sfrac{18}{35}
(\alpha Z) R 
\nonumber \\
& & - \sfrac{1}{3} R \left [ \pm 2 \overline{b} + \overline{d} \right ] ,
\nonumber \\
B_3^{GT} & \simeq & - \sfrac{4}{9} R^2 \left ( 1 - \sfrac{1}{10} x \right ) ,
\label{BnGT}
\eea
where
\be
x = - \sqrt{10} {\cal M}_{1y} / {\cal M}_{\sigma r^2} ,
\label{xxx}
\ee
\be
\overline{b}= \frac {1}{M R} \left [ \frac{g_M}{g_A} + \frac{{\cal M}_L}{{\cal M}_{GT}}
\right ] ,
\label{bbar}
\ee
\be
\overline{d} = \frac{1}{MR} \frac{{\cal M}_{\sigma L}}{{\cal M}_{GT}} ,
\label{dbar}
\ee
and also $\beta \simeq 6/5$, $g_M = 4.706$ and $M$ is the nucleon mass in electron rest-mass units.  
Where there is a $\pm$ symbol, the upper sign is used for electron emission beta decays, 
the lower sign for positron emitters.  All the transitions discussed in this work
are positron emitters, so the lower sign is consistently used.
The nuclear matrix
elements are defined in Eq.~(68) of Ref.~\cite{Hol74}.  Schematically, they are written:  ${\cal M}_{GT} = \langle \sigma 
\rangle$, ${\cal M}_{\sigma r^2} = \langle r^2 \sigma \rangle$, ${\cal M}_{1 y} = (16 \pi /5 )^{1/2} \langle r^2 \left
[ Y_2 \times \sigma \right ] \rangle$, ${\cal M}_L = \langle L \rangle$ and ${\cal M}_{\sigma L} = \langle \sigma \times L
\rangle $.  Note that the matrix element ${\cal M}_{\sigma L}$, and hence $\overline{d}$, vanishes in diagonal matrix
elements, as would occur in a mirror transition between isobaric analogue states.

The correction $\delta_S$ is again parameterized as in Eq.~(\ref{ds2}) with approximate expressions for the coefficients
derived from Eq.~(\ref{BnGT}).  For pure Gamow-Teller transitions they yield
\bea
b_0^{GT} & \simeq & \sfrac{2}{15} R^2 + \sfrac{1}{15} R^2 x
+ \sfrac{1}{3} \beta (\alpha Z) \left [ \pm 2 \overline{b} +
\overline{d} \right ] + \sfrac{61}{630} (\alpha Z)^2 ,
\nonumber \\
b_1^{GT} & \simeq & - \sfrac{26}{35}(\alpha Z) R - \sfrac{1}{35} (\alpha Z)
R x + \sfrac{1}{3} R \overline{d} - \sfrac{3}{2 M_A} ,
\nonumber \\
b_2^{GT} & \simeq & - \sfrac{9}{7} (\alpha Z) R - \sfrac{5}{6} R
\left [ \pm 2 \overline{b} + \overline{d} \right ] ,
\nonumber \\
b_3^{GT} & \simeq & - \sfrac{11}{105} R^2 \left ( 1 + \sfrac{1}{11} x \right ) .
\label{bbnGT}
\eea
Again in fitting the exact values of $\delta_S$ from Eq.~(\ref{dels}) with the expression in Eq.~(\ref{ds2}) we held the
parameters $b_2$ and $b_3$ fixed at the values given for $b_2^{GT}$ and $b_3^{GT}$ in Eq.~(\ref{bbnGT}) and then obtained
the parameters $b_0$ and $b_1$ from the fit.

The $T$=$\,$1/2 mirror transitions are mixed transitions, with both Fermi and Gamow-Teller components.  The fitted $b$ coefficients
for both the Fermi and Gamow-Teller components are given in Table~\ref{t:tab2} along with the $a_0$ and $a_1$ coefficients.  In such
mixed transitions the inverse of the partial lifetime is proportional to
\be 
t^{-1} \propto f_V \left [ |{\cal M}_F|^2 + \frac{f_A}{f_V} | g_A {\cal M}_{GT} |^2
\right ] ,
\label{invt}
\ee
where
\bea
f_V & = & f_0 \left ( 1 + \delta_S^F \right ) ,
\nonumber \\
f_A & = & f_0 \left ( 1 + \delta_S^{GT} \right ) .
\label{fVfA}
\eea
The statistical rate functions $f_V$ and $f_A$ are easily obtained from the parameters listed in Table~\ref{t:tab2}.

\section{Conclusions}
\label{Conc}

We have provided simple parameterizations of the statistical rate functions, $f$, for nuclear $\beta$ transitions of current interest
in determining $V_{ud}$ and testing CKM unitarity.  In most but not all cases, the transition $Q_{EC}$ values have already
been measured with $\sim$1-keV precision or better.  In a few cases they are much less well known.  In all cases, the $Q_{EC}$ values
will undoubtedly be remeasured, leading possibly to different values and certainly to reduced uncertainties.  When this happens,
experimenters will need $f$ values of equivalent precision, and the parameterizations presented here will satisfy that need without
complicated computing.

It is important to note that our parameterization is only valid for the transitions identified and only for a limited range of energies
($\pm$60 keV for all cases except for the decay of $^{70}$Br which covers $\pm$600 keV) around the currently accepted $Q_{EC}$ values for
those transitions \cite{HT15,Se08}.  The coefficients of our parameterization should not be applied outside the range of energies specified
or to any other transitions. 

\appendix
\section{Kinematic recoil corrections}
\label{s:recoil}

Let $M_A$ be the average mass of the initial and final nuclei.  Then the kinematic recoil corrections are of order $W_0/M_A$
and, in all but the most precise work, they can generally be ignored.  The recoil correction enters the calculation in two places:  
firstly, the end-point energy is slightly modified, a correction we denote $\Delta f^a$; and secondly, additional terms are added to
the shape-correction function $f_1(W)$, providing a correction we call $\Delta f^b$.

For the first correction: If $W_0$ is the end-point energy without consideration of recoil and $W_0^{\rm corr}$ is the corrected value, 
then from Eq.(3) of Holstein \cite{Hol74} we get
\bea
W_0^{\rm corr} & = & W_0 \left ( 1 + \frac{1}{2 W_0 M_A} \right )
\left ( 1 + \frac{W_0}{2 M_A} \right )^{-1}
\nonumber \\[1mm]
& \simeq & W_0 \left ( 1 - \frac{W_0}{2 M_A} + \frac{1}{2 W_0 M_A} \right ) .
\label{W0corr}
\eea
So, since the statistical rate function is approximately proportional to $W_0^5$, the correction to $f$ must be of order
\be
\frac{\Delta f^a}{f} \simeq 1 - \frac{5}{2} \frac{W_0}{M_A} + 
\frac{5}{2} \frac{1}{W_0 M_A}.
\label{Df1}
\ee

Unlike $\Delta f^a$, the recoil correction to the shape-correction function, $\Delta f^b$, is different for Fermi and Gamow-Teller
transitions.  The modifications are
\bea
f_1^{F,{\rm corr}}(W) & = & f_1^F(W) \left ( 1 + 2 \frac{W}{M_A} \right ) ,
\nonumber \\[2mm]
f_1^{GT,{\rm corr}}(W) & = & f_1^{GT}(W) \left ( 1 - \frac{2}{3} \frac{W_0}{M_A}
+ \frac{10}{3} \frac{W}{M_A} \right. 
\nonumber \\
& & ~~~~~~~~~~~~~~~~ \left. - \frac{2}{3} \frac{1}{M_A W} \right ).
\label{f1corr}
\eea
If these corrections are integrated over the electron spectrum, they yield corrections to the statistical rate function of
\bea
\frac{\Delta f^{b,F}}{f} & \simeq & 1 + \frac{W_0}{M_A} ,
\nonumber \\[1mm]
\frac{\Delta f^{b,GT}}{f} & \simeq & 1 - \frac{2}{3} \frac{W_0}{M_A}
+ \frac{5}{3} \frac{W_0}{M_A} - \frac{5}{3} \frac{1}{M_A W_0} .
\label{Df2}
\eea
Finally, combining corrections $\Delta f^a$ and $\Delta f^b$, we obtain the final recoil correction to the statistical rate function
\bea
\frac{\Delta f^F}{f} & = & 1 - \frac{3}{2} \frac{W_0}{M_A}
+ \frac{5}{2} \frac{1}{W_0 M_A} 
\nonumber \\
& \simeq & 1 - \frac{3}{2} \frac{W_0}{M_A} ,
\nonumber \\[1mm]
\frac{\Delta f^{GT}}{f} & = & 1 - \frac{3}{2}\frac{W_0}{M_A}
+ \frac{5}{6} \frac{1}{W_0 M_A} 
\nonumber \\
& \simeq & 1 - \frac{3}{2} \frac{W_0}{M_A}.
\label{Dffinal}
\eea
Thus, Fermi and Gamow-Teller transitions are subject to essentially the same correction and it is this correction that we have
recorded in Eq.~(\ref{recoil}) and used in our fitting algorithms.  Of course, the exactly computed $f$ values, to which our
parameterizations are fitted, include the complete kinematic recoil treatment.

\begin{acknowledgments}

This material is based upon work supported by the U.S. Department of Energy, Office of Science, Office of Nuclear Physics, under
Award Number DE-FG03-93ER40773, and by the Robert A. Welch Foundation under Grant No.\,A-1397.

\end{acknowledgments}


\begin{thebibliography}{99999}

\bibitem{HT09}
J.C. Hardy and I.S. Towner, \pr C {\bf 79}, 055502 (2009).

\bibitem{HT15}
J.C. Hardy and I.S. Towner, arXiv 1411.5987 and to be published.

\bibitem{Se08}
N. Severijns, M. Tandecki, T. Phalet and I.S. Towner, \pr C {\bf 78}, 055501 (2008).

\bibitem{HT05}
J.C. Hardy and I.S. Towner, \prc {\bf 71}, 055501 (2005) and \prl {\bf 94}, 092502 (2005).

\bibitem{Hol74}
B.R. Holstein, Rev. Mod. Phys. {\bf 46}, 789 (1974);
erratum, Rev. Mod. Phys. {\bf 48}, 673 (1976).

\bibitem{Ro36}
M.E. Rose, Phys. Rev. {\bf 49}, 727 (1936).



\bibitem{Sc66}
H. Schopper, {\it Weak Interactions and Nuclear Beta Decay}
(North-Holland, Amsterdam, 1966).

\bibitem{BB82}
H. Behrens and W. B\"{u}hring, {\it Electron Radial Wave Functions and 
Nuclear Beta-decay} (Clarendon Press, Oxford, 1982);
H. Behrens, H. Genz, M. Conze, H. Feldmeir, W. Stock and A. Richter,
Ann. of Phys. {\bf 115}, 276 (1978).

\end{thebibliography}
\end{document}